\documentclass[aip,reprint]{revtex4-1}
\usepackage{graphicx}
\usepackage{listings}
\usepackage{picins}
\usepackage{latexsym}

\draft 

\begin{document}


\title{Designing ABO$_{3}$ structure with Lennard-Jones interatomic potentials} 



\author{Hui Zhang}
\email[Electronic mail:]{zhope@scut.edu.cn}
\noaffiliation
\author{Zhongwu Liu}
\noaffiliation
\author{Xichun Zhong}
\noaffiliation
\author{Dongling Jiao}
\noaffiliation
\author{Wanqi Qiu}
\noaffiliation
\affiliation{School of Materials Science and Engineering, South China University of Technology, Guangzhou 510640, People's  Republic of China}


\date{\today}

\begin{abstract}
In this paper, our goal is to design ABO$_{3}$ crystal structure with simple interatomic Lennard-Jones (LJ) potentials and without setting any initial Bravais lattice and it is carried out by molecular dynamics (MD) simulation. In the simulation, the equilibrium distances between atoms are determined by LJ potentials. For the identification of the microstructure of simulated system, we have calculated the distribution functions of both the angles between one atom and its nearest neighbors and the distances between atoms and compared the results with those of ideal lattices. The results have clearly shown that we have successfully produced ABO$_{3}$ crystal structure by MD simulation.
\end{abstract}


\maketitle 

\section{Introduction}
Designing specific materials as needed is the goal of material scientists. However, this job is very difficult and more work is needed. In our previous work, it has been found that the interatomic potentials are very important for the formation of crystalline solids \cite{Zhang-1,Zhang-2}. Very complex crystal structures such as diamond and graphite structures can be formed with simple LJ interatomic potentials. This means that it is possible for us to design the system with desirable crystal structure even though we have not learnt a lot about the interaction between atoms. In the simulation of the crystal structures reported in Refs. 1 and 2, both the interactions and the equilibrium distances between atoms are defined by LJ potentials. One question of whether we can design the crystal structures of material by using the equilibrium distances determined by LJ potentials arises. To answer this question, we can choose one well-known and complex structure as a target, design the crystal structure in terms of its lattice constants, and reproduce it by MD simulation.  
ABO$_{3}$ structure, also called perovskite structure, is a very important crystal structure for many materials showing different functions such as ferroelectric crystals of BaTiO$_{3}$ and PbTiO$_{3}$ \cite{Rabe-3}, and perovskite solar cells \cite{Lotsch-4, Mitzi-5}. In ABO$_{3}$ structure, A atoms occupy the vortices of the cubic, B atoms the body center positions, and O atoms six face centers. We have taken AB$_{3}$ structure as our target. In our strategy, the equilibrium distances between atoms are determined by the LJ potentials. If the lattice constant of ABO$_{3}$ structure is $a$, then $r_{AA}$=$a$, $r_{BB}=a$, $r_{OO}=0.707a$, $r_{AB}=0.866a$, $r_{AO}=0.707a$, and $r_{BO}=0.5a$. With the above potential parameters, it is expected that the liquid-crystalline phase transition of ABO$_{3}$ system can be reproduced by MD simulation and at crystalline state our simulated system shows a perfect ABO$_{3}$ crystal structure. We identify the crystal structure of simulated system by calculating the distribution functions of both the angles between one atom and its nearest neighbors and the distances between atoms for A-A, B-B, O-O, A-O, B-O, and A-B subsystems and checking the atomic arrangements.\\

 \begin{figure}[h t b p]
\centering
 \includegraphics[width=80mm]{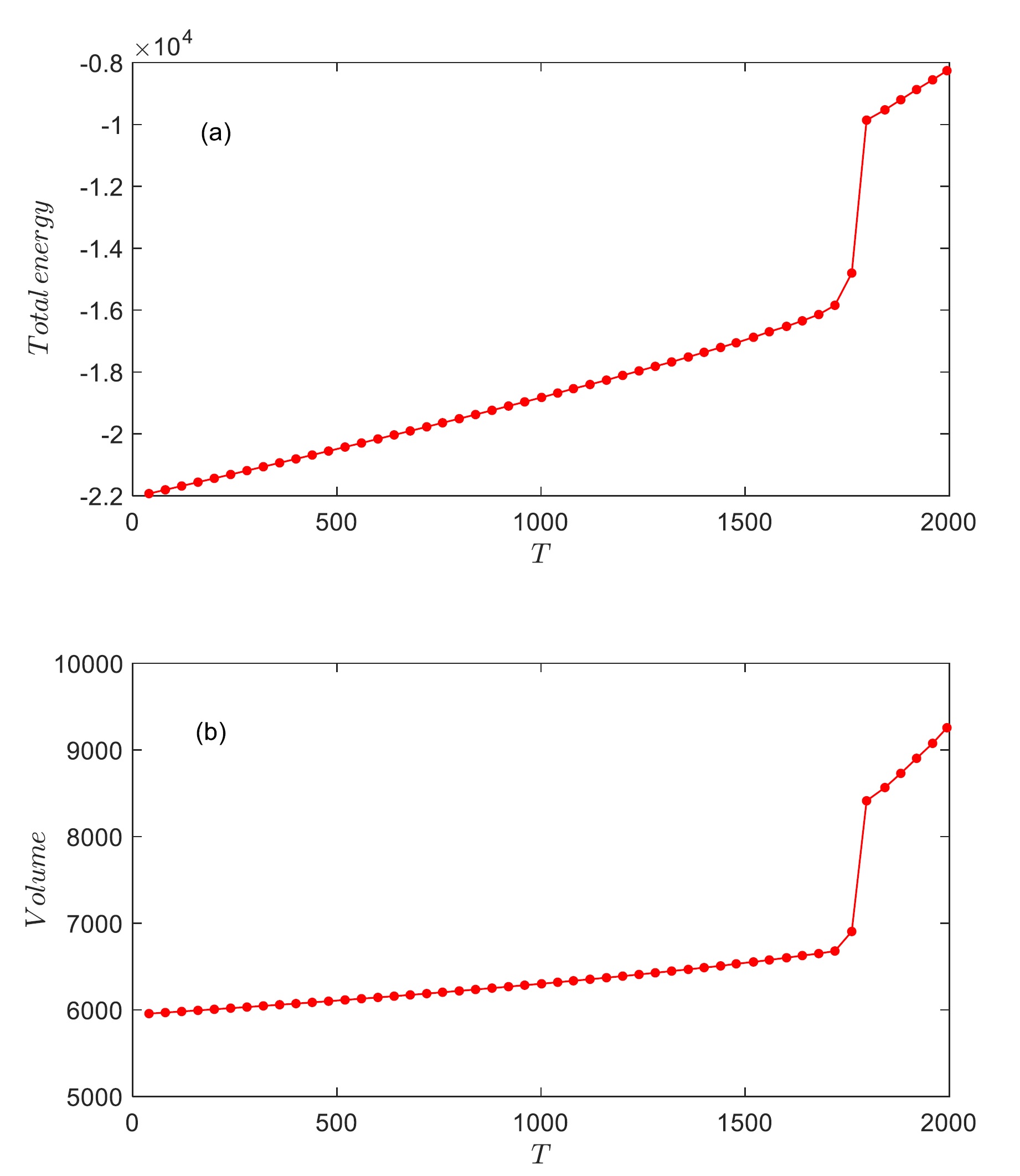}%
 \caption{\label{FIG.1.}Dependence of both the total energy (a) and volume (b) of simulated systems on the temperature.}
 \end{figure}
  \begin{figure*}
\centering
 \includegraphics[width=160mm]{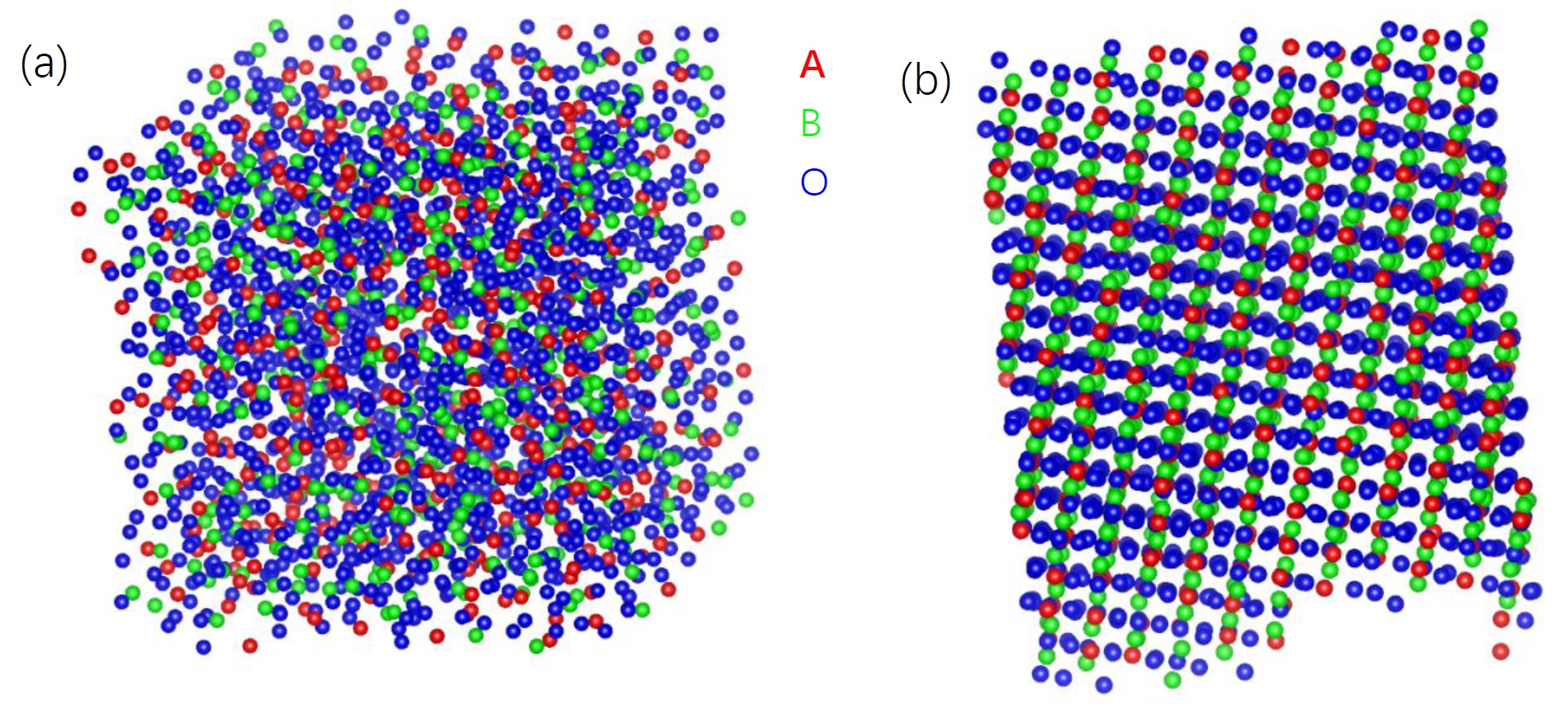}%
 \caption{\label{FIG.2.}The atomic arrangements of simulated system at the liquid state (a) and at the crystalline state (b).}
 \end{figure*}
 \begin{figure*}
\centering
 \includegraphics[width=160mm]{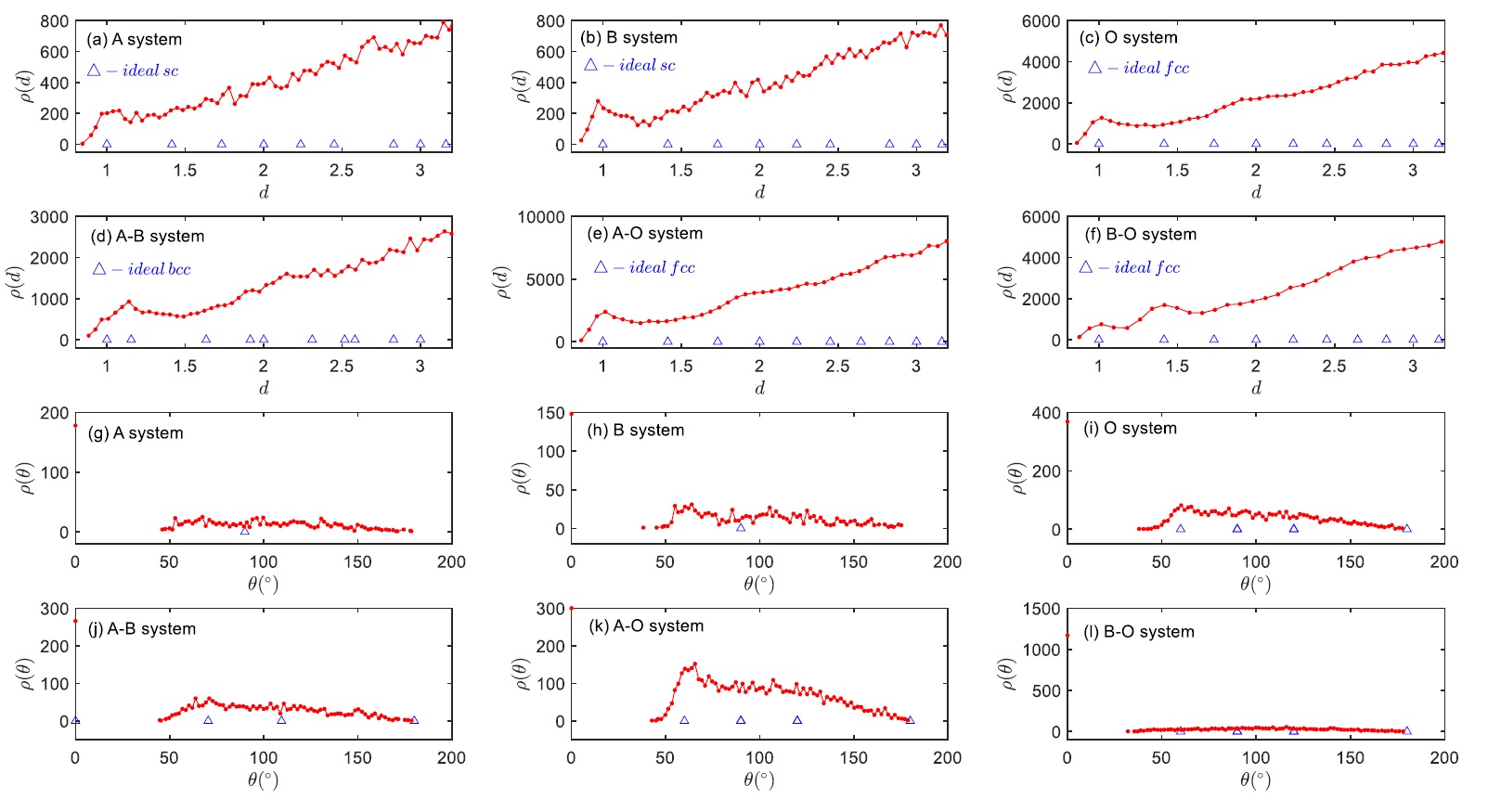}%
 \caption{\label{FIG.3.}At the liquid state, the distribution functions of both the angles between one atom and its nearest neighbors (a) and the distances between the atoms (b). ‘$\triangle$’ denote the distribution functions for ideal lattices.}
 \end{figure*}
\begin{figure*}
\centering
 \includegraphics[width=160mm]{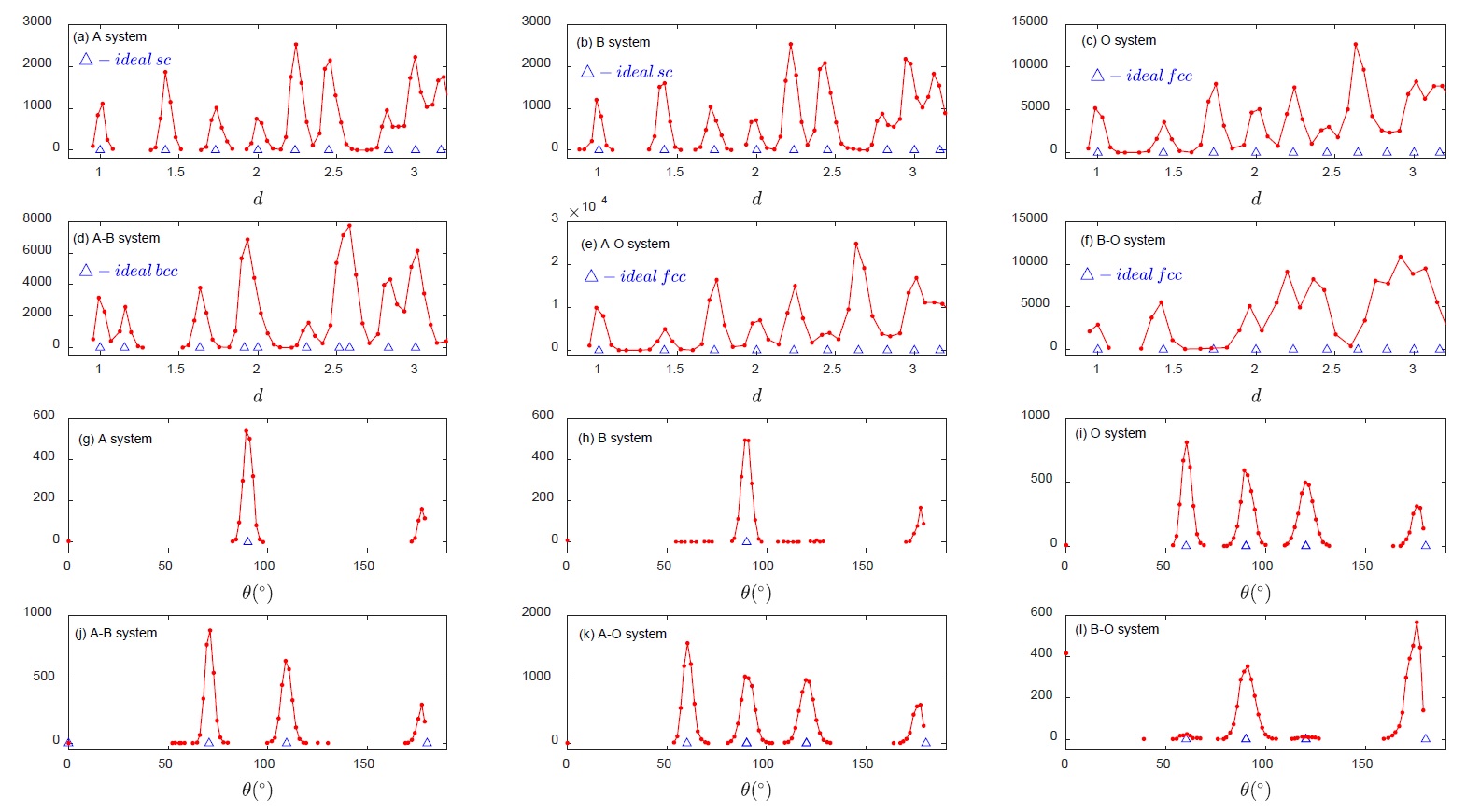}%
 \caption{\label{FIG.4.}At the crystalline state, the distribution functions of both the angles between one atom and its nearest neighbors (a) and the distances between the atoms (b). ‘$\triangle$’ denote the distribution functions for ideal lattices.}
 \end{figure*}
   \begin{figure*}
\centering
 \includegraphics[width=160mm]{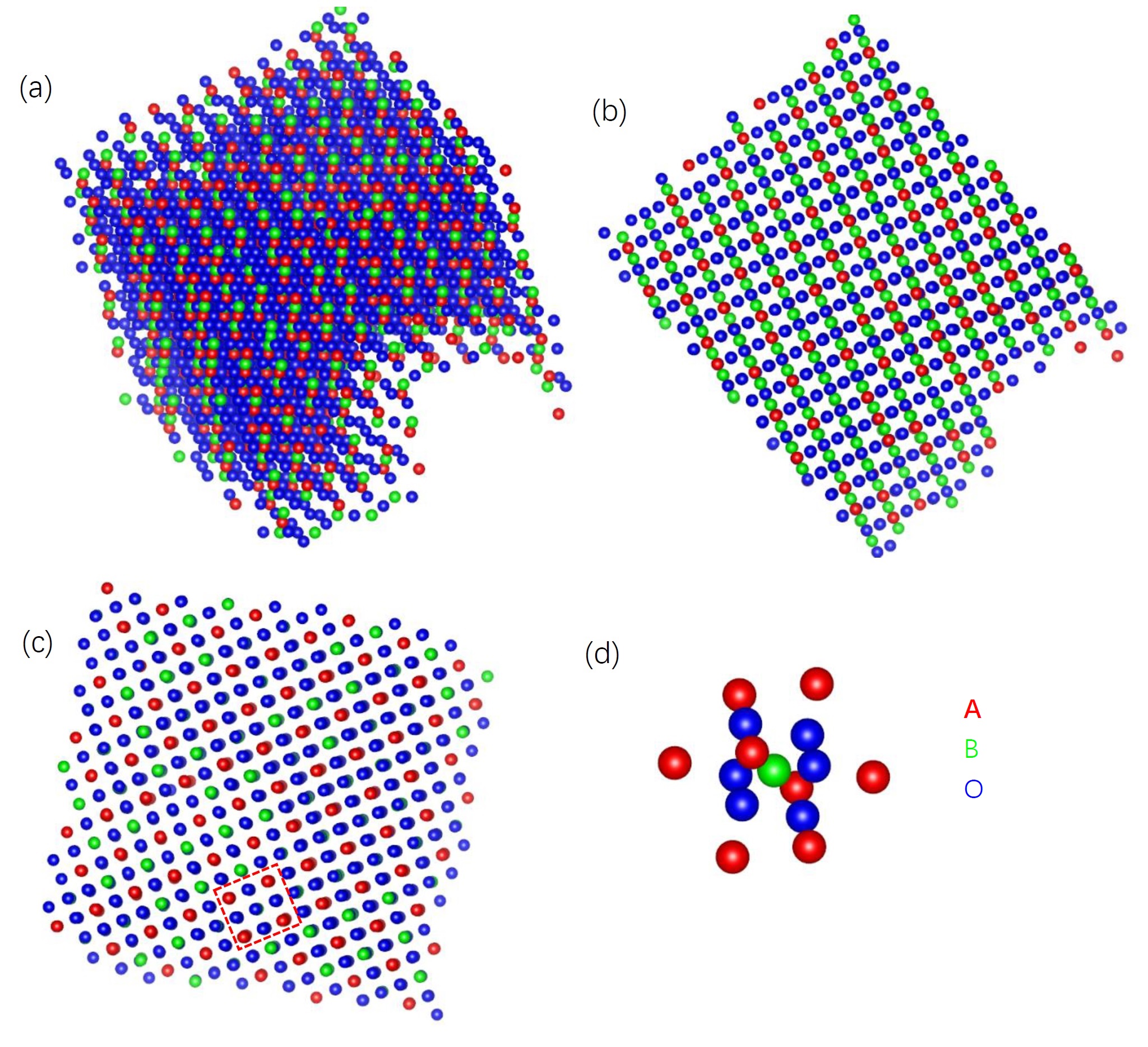}%
 \caption{\label{FIG.5.}(a)-(c)the atomic arrangements of simulated system observed from different directions, (d) ABO$_{3}$ crystal cell obtained from the region encircled by the red dashed lines shown in (c).}
 \end{figure*}
 
\section{Simulation and method}
We introduce the classic LJ potential to describe the interatomic coupling, and LJ potential can be written as
\begin{eqnarray}
U(r)=4\epsilon \left(\left(\frac{\sigma}{r}\right)^{12}-\left(\frac{\sigma}{r}\right)^{6}\right)
\end{eqnarray}
where $\epsilon$ is the depth of the potential well, $\sigma$ is the finite distance at which the inter-particle potential is zero, and $r$ is the distance between particles. The distance of cutoff is denoted by $r_{c}$. The simulation was carried out with the aid of LAMMPS \cite{Plimpton-6}. In the simulation, the LJ units were used, and the periodic boundary conditions were applied. We set $r_{c}$=3$\sigma$. $\sigma_{AA}$=2.0, and $r_{c}$=6.0. $\sigma_{BB}$=2.0, and $r_{c}$=6.0. $\sigma_{OO}$=1.414, and $r_{c}$=4.242. $\sigma_{AB}$=1.732, and $r_{c}$=5.196. $\sigma_{AO}$=1.414, and $r_{c}$=4.242. $\sigma_{BO}$=1.0, and $r_{c}$=3.0. The masses of the particles were all 100, and the numbers of particles were denoted by $n$. $n_{A}$=500, $n_{B}$=500, and $n_{O}$=1500. We did not set any initial Bravais lattice. The particles were created randomly in the simulation box and then an energy minimization procedure followed. The initial temperature $T_{0}$=$\epsilon$, and $\epsilon$=2000. We set the timestep as 0.001. At $T_{0}$, NPT dynamics was implemented for 1$\times$10$^{6}$ timesteps, and then the temperature was decreased by $T_{0}$/40. At every following temperatures $T$, NPT was carried out for a time of $t$ timesteps, and $t$=10$^{6}$. The pressure was always zero in the simulation. Details can be found in in-script in Appendix. The visualization of simulated results was done with the aid of VESTA \cite{Momma-7}. For LJ potential, the equilibrium distance $r_{0}$=1.12$\sigma$.\\
\indent We calculated the distribution functions of both the angles between one atom and its nearest neighbors and the distances between atoms for the identification of microstructure of simulated system\cite{Zhang-2}. For more information, a set of A (or B and O) atoms was treated as a subsystem, and called A, B, and O systems. A-B system was set as a union of both A subsystem and B subsystem, and we had A-B, A-O, and B-O systems.\\

\section{Results and Discussions}
When ABO$_{3}$ system is cooled from high temperature, there is a liquid-solid phase transition, and an abnormal jump in its volume. Figure 1 shows the dependence of both the total energy and volume of simulated system on the temperature. As shown in Fig. 1, sudden changes in the total energy and volume clearly show that there is a phase transition. This means that there is a disorder-order phase transition and an unusual change in the atomic arrangement. Figure 2 shows the atomic arrangements at the liquid and crystalline states. As shown in Fig. 2, the difference in the atomic arrangements is clear but we cannot determine which lattice the system shows. In order to identify the crystal structure, we calculated the distribution functions of both the angles between one atom and its nearest neighbors and the distances between atoms for A, B, O, A-B, A-O, and B-O systems. When their distribution functions are in agreement with those of ideal lattices, we can make the final identification of the crystal structure. Figure 3 shows the distribution functions of both the angles between one atom and its nearest neighbors and the distances between atoms. As shown in Fig. 3, there are no clear peaks associate with a specific lattice, and our simulated system is in a liquid state. Figure 4 shows the distribution functions at crystallization state. In Fig. 4(a), A atoms form a simple cubic (sc) lattice, and the angles between its nearest neighbors are 90$^\circ$. The ratios of the distances between atoms $d_{1}$: $d_{2}$: $d_{3}$: $\cdots$=1: 1.414: 1.732: $\cdots$, and the results are in agreement with data of ideal sc lattice, indicating that the atomic arrangements of A atoms are correct. In Fig. 4(b), B atoms occupy the body center of ABO$_{3}$ structure, and all B atoms also form a sc lattice. In Fig. 4(c), O atoms occupy the face centers of ABO$_{3}$ cubic, the angles between these O atoms are 60$^\circ$, 90$^\circ$, and 120$^\circ$. In Fig. 4(d), A and B atoms in ABO$_{3}$ structure form a body-centered cubic (bcc) lattice, and the angles between A and B atoms and its nearest neighbors are 70.5$^\circ$ and 109.5$^\circ$. In Fig. 4(e), A and O atoms form a face-centered cubic (fcc) lattice, and the angles between its nearest neighbors are 60$^\circ$, 90$^\circ$, and 120$^\circ$. In Fig. 4(f), B and O atoms form a tetrahedron, and the angles between B and O atoms are 90$^\circ$. It has been shown from Fig. 4 that the calculated results are in agreement with those of ideal lattices and the atomic arrangement of A, B, and O atoms are correct. Therefore, we can initially identify the system showing ABO$_{3}$ structure, and we further check the atomic arrangement for the final identification.\\
\indent Figure 5 shows the atomic arrangements of simulated system observed from different directions. As shown in Figs. 5(a)-(c), our simulated system is ordered, but we cannot make the final identification from these patterns. In Fig. 5(c), we tried to retrieve one ABO$_{3}$ crystal cell. We removed the atoms outside the region encircled by the red dash lines and rotated the residual structure. The same operation was repeated, and we obtained the ABO$_{3}$ crystal cell shown in Fig. 5(d). Till now, we finally identify the system showing the ABO$_{3}$ structure.\\
\indent It must be pointed out that we introduced in only the equilibrium distances between atoms and both the electric charge and electric spin are not involved.
\section{Conclusions}
Without setting any initial Bravais lattice and with simple Lennard-Jones interatomic potentials, simulated system showing ABO3 structure has been produced by MD simulation. LJ potentials presented can not only describe the liquid-crystalline phase transition but also determine the crystal structure of simulated system.
%
%
%
\appendix
\section{in script}
\begin{lstlisting}
units         lj
boundary      p p p
atom_style    atomic
dimension     3
region        box block 0 20 0 20 0 20
create_box    3 box
create_atoms  1 random 500 245 box
create_atoms  2 random 500 255 box
create_atoms  3 random 1500 265 box
timestep      0.001
thermo        1000
group         big1 type 1
group         big2 type 2
group         big3 type 3
mass          1 100
mass          2 100
mass          3 100
variable      lcut1 equal 3.0
variable      lcut2 equal 4.242
variable      lcut3 equal 5.196
variable      lcut4 equal 6.0
variable      la equal 2000
pair_style    lj/cut ${lcut1}
pair_coeff    1 1 ${la} 2.0 ${lcut4}
pair_coeff    2 2 ${la} 2.0 ${lcut4}
pair_coeff    3 3 ${la} 1.414 ${lcut2}
pair_coeff    1 2 ${la} 1.732 ${lcut3}
pair_coeff    1 3 ${la} 1.414 ${lcut2}
pair_coeff    2 3 ${la} 1.0 ${lcut1}

minimize      1.0e-10 1.0e-10 1000000 &
              1000000
dump          1 all image 1000 image.* &
              .jpg type type zoom 1.6
run           1000
undump        1
variable      ltemp equal 1600
velocity      all create ${ltemp} &
              314029 loop geom
fix           1 all npt temp ${ltemp} &
              ${ltemp} 1.0 iso 0.0 &
              0.0 1.0
dump          1 all image 500000 image &
              .*.jpg type type zoom 1.6
dump          2 all xyz 500000 file1.&
              *.xyz
run           1000000
undump        1
undump        2
variable      lpa equal ${ltemp}
label         loopa
variable      a loop 39
variable      tem equal ${lpa}-40*$a
fix           1 all npt temp ${tem} &
              ${tem} 1.0 iso 0.0 0.0 1.0
dump          1 all image 5000 image.* &
              .jpg type type zoom 1.6
dump          2 all xyz 5000 file1.*.xyz
run           1000000
undump        1
undump        2
next          a
jump          SELF loopa



\end{lstlisting}
\begin{acknowledgments}
This work is supported by the National Natural Science Foundation of China (Grant No. 11204087).
\end{acknowledgments}

\end{document}